\documentclass[10pt,letterpaper]{article}
\usepackage{opex3}
\usepackage{color}
\usepackage{units}
\usepackage{amsmath}
\usepackage{amssymb}
\usepackage{graphicx}
\usepackage{bm}
\usepackage{multirow,color,relsize,ulem,microtype,cite}

\newcommand{\be}{\begin{equation}}
\newcommand{\ee}{\end{equation}}

\newcommand{\gper}{\gamma_\perp}
\newcommand{\Eq}[1]{Eq.~(\ref{#1})}

\begin{document}

%%%%%%%%%%%%%%%%%% title page information %%%%%%%%%%%%%%%%%%
\title{Interaction-induced mode switching in steady-state microlasers}

\author{Li Ge,$^{1,2,*}$ David Liu,$^3$ Alexander Cerjan,$^4$ Stefan Rotter,$^5$ Hui Cao,$^4$ Steven G. Johnson,$^6$ Hakan E. T\"ureci,$^7$ and A. Douglas Stone$^4$}

\address{$^1$Department of Engineering Science and Physics, College of Staten Island, CUNY, Staten Island, NY 10314, USA \\
$^2$The Graduate Center, CUNY, New York, NY 10016, USA \\
$^3$Department of Physics, Massachusetts Institute of Technology, Cambridge, MA 02139, USA \\
$^4$Department of Applied Physics, Yale University, New Haven, CT 06520, USA\\
$^5$Institute for Theoretical Physics, Vienna University of Technology, A-1040 Vienna, Austria, EU \\
$^6$\textls[-18]{Department of Mathematics, Massachusetts Institute of Technology, Cambridge, MA 02139, USA} \\
$^7$Department of Electrical Engineering, Princeton University, Princeton, NJ 08544, USA}

\email{$^*$li.ge@csi.cuny.edu} %% email address is required

\begin{abstract}
We demonstrate that due to strong modal interactions through cross-gain saturation, the onset of a new lasing mode can switch off an existing mode via a negative power slope. In this process of interaction-induced mode switching (IMS) the two involved modes maintain their identities, i.e. they do not change their spatial field patterns or lasing frequencies. For a fixed pump profile, a simple analytic criterion for the occurrence of IMS is given in terms of their self- and cross-interaction coefficients and non-interacting thresholds, which is verified for the example of a two-dimensional microdisk laser. When the spatial pump profile is varied as the pump power is increased, IMS can be induced even when it would not occur with a fixed pump profile, as we show for two coupled laser cavities. Our findings apply to steady-state lasing and are hence different from dynamical mode switching or hopping. IMS may have potential applications in robust and flexible all-optical switching.
\end{abstract}

\ocis{(140.3430) Laser theory; (140.3945) Microcavities.}

%%%%%%%%%%%%%%%%%%%%%%% References %%%%%%%%%%%%%%%%%%%%%%%%%

%%%%%%%%%%%%%%%%%%%%%%%%%%  body  %%%%%%%%%%%%%%%%%%%%%%%%%%
\section{Introduction}
Our understanding of lasers is largely based on the semiclassical laser theory, which describes successfully modal thresholds, output power, spatial mode patterns inside and outside of the laser cavity, as well as dynamical effects such as relaxation oscillations and mode, phase, and frequency locking \cite{Haken,Scully}. With the advent of microlasers \cite{Microcavity1,Microcavity2}, their non-Hermitian nature and strong modal interactions not only introduced new phenomena (see, for example, \cite{EP,Science}) but also imposed new challenges for the understanding of these novel lasers. A new framework based on the semiclassical laser theory, known as the Steady-state Ab-initio Laser Theory (SALT) \cite{Science,TS,TSG,SPASALT,C-SALT}, addresses these challenges satisfactorily when the level inversion in the gain medium can be treated as stationary. SALT has been shown to give excellent agreement with much more computationally intensive  finite-difference-time-domain (FDTD) simulations of the lasing equations \cite{OpEx08,N_level,directMethod}, and it also accurately predicts recent experimental observations of exceptional points in lasers \cite{EP,EP_exp}. These advantages make SALT an ideal tool to study the effects of non-linear modal interactions in the steady-state, in regimes not previously studied and in some cases not accessible with other approaches.

\begin{figure}[t]
\centering
\includegraphics[clip,width=0.7\linewidth]{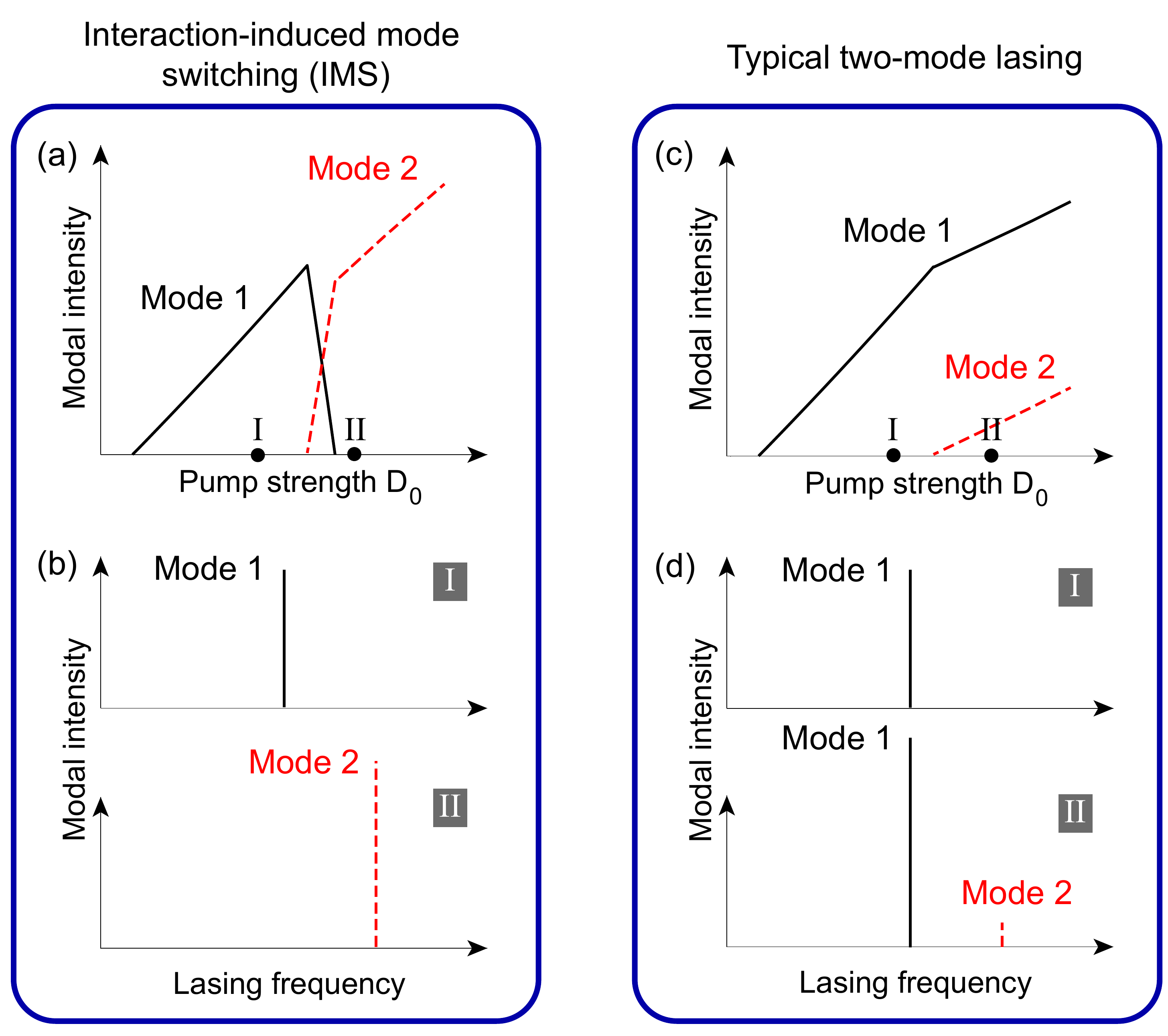}
\caption{Schematics showing modal intensities as a function of the pump strength for (a) interaction-induced mode switching and (c) typical two-mode lasing. (b,d) The corresponding lasing spectra at pump strengths marked by I and II in (a,b).}\label{fig:schematics}
\end{figure}

In this report, we reveal a surprising phenomenon due to such modal interactions: the onset of one lasing mode can switch off another mode which is already lasing %under certain circumstances
[Figs.~\ref{fig:schematics}(a) and \ref{fig:schematics}(b)]; we refer to this phenomenon as interaction-induced mode switching (IMS). IMS is very different from typical multimode lasing, where the modal intensity of each mode increases linearly with the pump strength and the power slopes of lasing modes are only slightly reduced at the onset of a new mode [Figs.~\ref{fig:schematics}(c) and \ref{fig:schematics}(d)]. %While mode switching due to non-linear interactions has previously been seen in SALT simulations of two-dimensional random lasers \cite{Science} in the quasi-ballistic limit,
IMS is a robust phenomenon and can in principle occur in any type of laser, including standard systems such as a microdisk and coupled one-dimensional cavities.
We note that IMS is a \textit{steady-state} effect and different from dynamical mode switching and related phenomena \cite{Mandel_book,Kawaguchi}. It is also independent of laser bistability \cite{Lamb, bistability1,bistability2, bistability3}, which can also lead to various mode switching scenarios. In addition, IMS differs qualitatively from seemingly similar effects found in our previous work \cite{EP,Science}, as we will explain in detail below.

In the next section we first review the basic ansatz and results of SALT and its ``single-pole" approximation (SPA-SALT), where each lasing mode can be well approximated by a linear mode, the constant-flux (CF) state \cite{TS,SPASALT,ge_NP}. We then present an example of IMS in a uniformly pumped two-dimensional (2D) microdisk laser. The presence of IMS in this example can be predicted and understood via a simple analytic criterion obtained from SPA-SALT, involving only a few parameters, i.e., the self- and cross-gain saturation coefficients of the two modes (``interaction coefficients") and the thresholds of the modes in the absence of gain saturation (``non-interacting thresholds").
The prediction of IMS from SPA-SALT is confirmed by the use of the full SALT computation.
Next, we show that IMS can also be facilitated by having a spatial pump profile which evolves as the overall pump power is increased. An example is given using two coupled one-dimensional (1D) slab cavities, in which IMS does not occur when the cavities are pumped uniformly. With different choices of evolving the spatial pump profile, we show that either of two modes can undergo an IMS, i.e. mode 1 be switched off by mode 2 or vice-versa, indicating the possibility of flexible control of the mode spectrum by using IMS and variable pump profile. A criterion for IMS in this case is again given using SPA-SALT, which shows an invariant property of IMS and its less restrictive requirement on the interaction coefficients.

\section{SALT and SPA-SALT}

SALT makes a steady-state ansatz and reduces the time-dependent coupled lasing equations for the electric field, level populations and polarization of the gain medium, to a set of non-linear time-independent wave equations for the lasing modes and their frequencies
\cite{Science,TS,TSG,SPASALT,C-SALT,OpEx08,N_level}.  In the case of multimode lasing it takes advantage of a separation of time scales between the slow population dynamics and the beat frequencies of the modes to predict a stable multimode stationary state, an approach which is typically valid only for microlasers \cite{directMethod,HS2,SA}.
While SALT neglects the time-dependent effects of the four-wave mixing non-linearity, it treats the time-independent effects of gain saturation and spatial hole-burning to infinite order and hence is found to be accurate far above threshold in steady-state lasing. In the discussion below, diffusion of excitations in the gain medium is assumed absent; recent work has shown how to include this effect in the SALT framework \cite{C-SALT}.

SALT differs from the standard modal description of semiclassical laser theory \cite{Haken,Scully} in a few important aspects: SALT describes the openness of the lasing medium rigorously in terms of an outgoing boundary condition, and, as noted, the static modal interaction is treated to infinite order and not truncated at the cubic level. The first method developed for solving the SALT equations and the one most useful in the current work is through
the expansion of a lasing mode $\Psi_\mu(\vec{r};D_0)$ at a given pump strength $D_0$ in an optimal basis of CF states $\{u_p(\vec{r};\omega_\mu)\}$, i.e. $\Psi_\mu(\vec{r};D_0) = \sum_p a^{(\mu)}_p(D_0)u_p(\vec{r};\omega_\mu)$.  Each CF state is purely outgoing and parametrized by the lasing frequency $\omega_\mu$, and the full set is biorthogonal and complete; hence an arbitrary lasing mode can be represented in this fashion.  %In addition, just above the first threshold, the lasing mode is given by a single (scaled) CF state, so that further above threshold the CF representation converges quite rapidly to the true solution. %% We use UCF in this paper, so this statement is not exactly true, especially for the case with evolving pump profile.
We define the modal intensity by $I_\mu(D_0) = \sum_p|a^{(\mu)}_p(D_0)|^2$, and its power slope, $S_\mu = d{I_\mu(D_0)}/dD_0$, is the key quantity of interest in the study of IMS.

In the previous expressions, the lasing frequency $\omega_\mu$ is a real quantity to be solved together with the mode profile $\Psi_\mu(\vec{r})$. The pump strength is represented as a dimensionless quantity $D_0$ proportional to the level inversion in the gain medium, and its spatial profile is assumed given by $f(\vec{r})$ and normalized by $\int_\text{cavity} f(\vec{r}) d\vec{r} =\int_\text{cavity} d\vec{r}\equiv V$.
The biorthogonality relation of the CF states can be shown to be \cite{TS}
\be
\int_\text{cavity} d\vec{r}\, \bar{u}^*_p(\vec{r})u_q(\vec{r}) = V\delta_{pq}\label{eq:biortho}
\ee
when the refractive index $n$ is uniform inside the cavity. $\bar{u}_p(\vec{r})$ is the biorthogonal partner of $u_p(\vec{r})$ and $``*"$ denotes complex conjugation (for a lossless cavity without degeneracy, the partner is simply $u_p(\vec{r})$ itself). In the next section we consider degenerate CF states, the traveling-wave (clockwise (CW) or counterclockwise (CCW)) modes in a microdisk, i.e.,
\be
u_p(r,\phi) \propto J_{m}\left(\frac{n\Omega_pr}{c}\right)e^{im \phi},~r < R.
\label{eq:CF2D1}
\ee
In this case $\bar{u}_p(\vec{r})$ is then given by %\cite{Ge_thesis}
\be
\bar{u}_p(r,\phi) \propto J_{m}\left(\frac{n^*\Omega_p^*r}{c}\right)e^{im \phi},~r < R.
\label{eq:CF2D2}
\ee
Here $\phi$ is the azimuthal angle, $J_m$ is the Bessel function of azimuthal quantum number $m$, $\Omega_\mu$ is the complex CF frequency, $c$ is the speed of light in vacuum, and $R$ is the radius of the microdisk cavity.

When the $Q$-factors of the lasing modes are relatively high, each lasing mode can be well approximated by a single CF state, i.e., there is only one significant $a^{(\mu)}_p$ in the expansion of $\Psi_\mu$, even when well above the laser threshold and when the pump profile $f(\vec{r})$ is non-uniform \cite{SPASALT,ge_NP}. This property enables us to invoke SPA-SALT, i.e., $\Psi_\mu(\vec{r};D_0) \approx a^{(\mu)}_p(D_0)u_p(\vec{r};\omega_\mu)$ and $I_\mu(D_0)\approx|a^{(\mu)}_p|^2$. This approximation neglects the power-dependence of the lasing frequencies and the change in the spatial mode profiles above threshold, but has the great advantage
of reducing the non-linear coupled SALT equations  to a set of (constrained) linear equations for the modal intensities
\cite{SPASALT}. For example, in the two-mode regime, the modal intensity $I_1$ and $I_2$ are determined by
\be
M
\begin{pmatrix}
I_1 \\ I_2
\end{pmatrix}
=
\begin{pmatrix}
\frac{D_0}{D_0^{(1)}} - 1 \vspace{4pt}\\
\frac{D_0}{D_0^{(2)}} - 1
\end{pmatrix},\quad
M
\equiv
\begin{pmatrix}
\Gamma_1\chi_{11} & \Gamma_2\chi_{12} \\
\Gamma_1\chi_{21} & \Gamma_2\chi_{22}
\end{pmatrix},\label{eq:SPASALT}
\ee
with the constraint that $I_1,I_2\geq0$.
Here $\Gamma_\mu\equiv  \gper^2/[\gper^2+(\omega_\mu-\omega_a)^2]\leq1$ is the Lorentzian gain factor for mode $\mu$, where $\omega_a$ is the atomic transition frequency and $\gper$ is the longitudinal relaxation rate of the gain medium. The self-interaction coefficients $\chi_{11},\chi_{22}$ and the cross-interaction coefficients $\chi_{12},\chi_{21}$ are given by
\be
\chi_{pq} = \frac{1}{V}\left|\int_\text{cavity} d\vec{r}\, \bar{u}^*_p(\vec{r})u_p(\vec{r}) \left|u_q(\vec{r})\right|^2\right|.\label{eq:chi}
\ee
We note that the product $n\Omega_\mu$ is almost real for a relatively high-$Q$ mode, even when the passive cavity refractive index $n$ has a small imaginary part due to the inclusion of material absorption. Therefore, using the expressions (\ref{eq:CF2D1}) and (\ref{eq:CF2D2}), the cross-interaction coefficients are approximately equal, i. e.,
\be
\chi_{12}\approx\chi_{21}\propto \int_\text{cavity} \hspace{-3pt}d\vec{r} \left|J_{m_1}\left(\frac{n\Omega_1r}{c}\right)J_{m_2}\left(\frac{n\Omega_2r}{c}\right)\right|^2
\ee
in the microdisk laser to be discussed in the next section.

Since we are using the microdisk laser as an example, we need to discuss the conditions for the validity of SALT (and SPA-SALT) in more detail. The rates which determine the laser population dynamics (in the simplest case of the Maxwell-Bloch model of two-level gain atoms) are described by a single parameter, denoted as
$\gamma_{\parallel}$.  Multimode SALT is approximately valid when $\gper,\Delta \gg \gamma_{\parallel}$ \cite{TS,OpEx08,directMethod,HS2}, where
$\Delta$ is the typical beat frequency between lasing modes (of order the free spectral range).  For the degenerate CW and CCW modes of a microdisk this condition is not satisfied, and we will assume that, as is typical in ring lasers, one sense of rotation is chosen by small perturbations, and SALT can then be used for the resulting non-degenerate lasing modes.  A more detailed treatment of the degenerate case can be found, for example, in \cite{SA}, but it is not relevant for the current discussion. We also note that the electric field and level inversion are measured in their respective natural units $e_c=\hbar\sqrt{\gper\gamma_\|}/2g$ and $d_c=\hbar\gper/4\pi g^2$, where $g$ is the dipole matrix element between the lasing transition levels.

The key quantity of interest in the discussion of IMS in the two-mode regime is the power slope of the first mode, as we have shown schematically in Fig.~\ref{fig:schematics}(a). Within SPA-SALT all lasing modes have a linear variation with pump power if the spatial pump profile is fixed,
featuring a slope which only changes when new modes turn on or off.
For the two mode case, it is straightforward to show that the power slope of the first mode is $S_1 = 1/[\Gamma_1\chi_{11}D_0^{(1)}]$ after it turns on at $D_0=D^{(1)}_0$, in which $\chi_{11}^{-1}$ plays the role of mode volume; this slope changes to
\be
\tilde{S}_1 =
% \frac{\chi_{22}/D_0^{(1)} - \chi_{12}/D_0^{(2)}}{\Gamma_1(\chi_{11}\chi_{22}-\chi_{12}\chi_{21})} =
\frac{\frac{\chi_{22}}{\chi_{12}}-\frac{D_0^{(1)}}{D_0^{(2)}}}{\frac{\chi_{22}}{\chi_{12}} - \frac{\chi_{21}}{\chi_{11}}} S_1 \label{eq:S}
\ee
once the second mode starts lasing at
\be
D_{0,\text{int}}^{(2)}
%= \frac{\chi_{11}-\chi_{21}}{\chi_{11}-\chi_{21}\frac{D_0^{(2)}}{D_0^{(1)}}}D_0^{(2)}.
 = \frac{1-\frac{\chi_{21}}{\chi_{11}}}{\frac{D_0^{(1)}}{D_0^{(2)}}-\frac{\chi_{21}}{\chi_{11}}}D_0^{(1)}.\label{eq:Dth2}
\ee
%In the absence of cross interaction, i.e., $\chi_{12}=\chi_{21}\rightarrow0$, we immediately find $\tilde{S}_1\rightarrow S_1$.
The subscript ``int'' here indicates that $D_{0,\text{int}}^{(2)}$ is the threshold of mode 2 in the presence of modal interaction with mode 1. In contrast, $D_0^{(2)}$ in Eqs.~(\ref{eq:S}) and (\ref{eq:Dth2}) is the threshold of mode 2 when the modal interaction is neglected, which is determined by the openness of the laser cavity and the material absorptions within, as well as the Lorentzian gain factor $\Gamma_\mu$ \cite{ge_NP}. We will refer to $D_0^{(\mu)}$ as the non-interacting threshold of mode $\Psi_\mu$. Since there is only one lasing mode at the first threshold, $D_{0,\text{int}}^{(1)}=D_0^{(1)}$ by definition.

Within SPA-SALT the full lasing behavior is captured by the set of interacting thresholds and power slopes between thresholds; these are determined solely by the non-interacting lasing frequencies and thresholds, the gain curve parameters and the interactions coefficients.  By definition a mode which has just turned on has a positive power slope; IMS occurs when at some interacting threshold the power slope of a previously lasing mode becomes negative, e.g., $\tilde{S}_1<0$ in the two-mode case above.  This criterion is completely general and can be applied to any laser system for which SPA-SALT is a good approximation. %When the spatial pump profile is varied with increasing pump power the lasing intensities do not vary linearly with pump and so a more complex criterion needs to be used, but this case can also be treated with SPA-SALT, and will be discussed in section ?
%The relative change of the power slope, $\tilde{S}_1/S_1$, is a dimensionless number and does not depend on the choice of the unit of the pump strength.

The SPA-SALT equations (\ref{eq:SPASALT}) and their multimode form may look similar to the Haken-Sauermann equations \cite{HS2,HS} and their generalizations (e.g., \cite{Deych}), but the latter employ the cubic approximation for the modal interactions and lead to unphysical mode behaviors for pump strength $\sim20\%$
above the first threshold \cite{Scully}. A detailed discussion of these points can be found in \cite{SPASALT,OpEx08}, where SALT is also compared to the work by Mandel and coworkers \cite{Mandel,Mandel2}.

\section{IMS with a fixed pump profile}

In this section we give an example of IMS in a 2D microdisk laser. We consider uniform pumping ($f(\vec{r})=1$) and transverse magnetic modes, with $\Psi_\mu(\vec{r})$ representing the out-of-plane electric field in the steady state.
We restrict our treatment to CCW traveling waves solutions for the reasons discussed previously.

\begin{figure}[t]
\centering
\includegraphics[clip,width=0.9\linewidth]{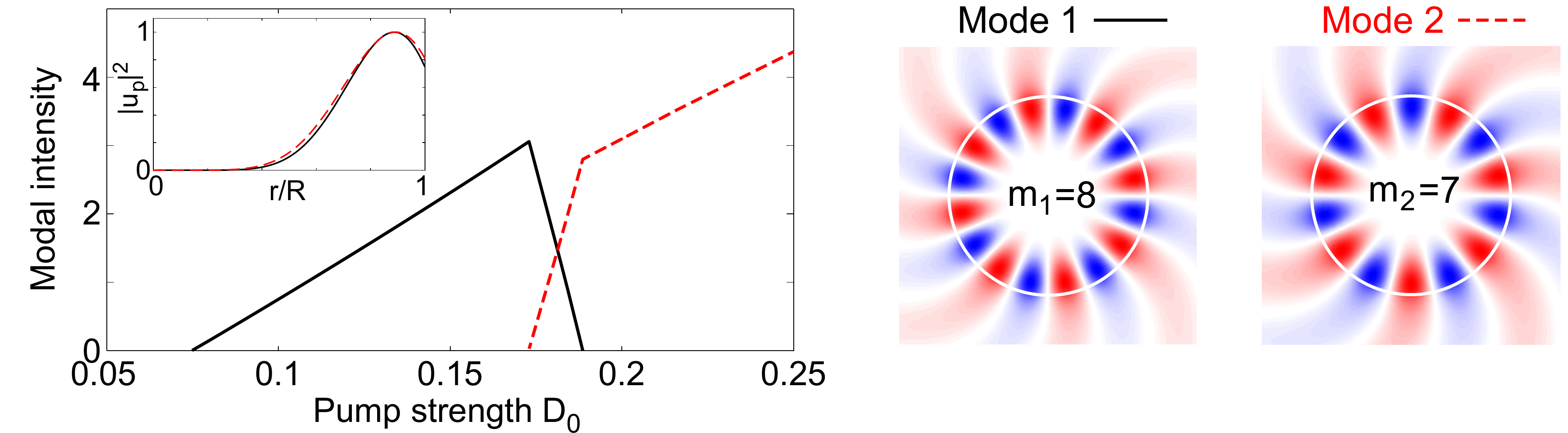}
\caption{Interaction-induced mode switching in a microdisk laser. Lower panel: Black solid and red dashed lines show the modal intensities of the first two modes. They are whispering-gallery modes with azimuthal quantum numbers $m_1=8$, $m_2=7$ and lasing frequencies $\omega_1R/c=5.37,\omega_2R/c=4.81$, respectively. Upper panel: False color plots of the real part of the electric field in these two modes, where red and blue indicate positive and negative values. Their radial profiles are shown as the inset in the lower panel.
The parameters used are: atomic transition frequency $\omega_aR/c=4.83$, longitudinal relaxation rate $\gper R/c=1$ and refractive index $n=2+0.01i$.}\label{fig:disk}
\end{figure}

Using the full SALT, we find that a whispering gallery mode of azimuthal quantum number $m_1=8$ starts lasing when the pump strength is higher than its threshold $D_0^{(1)}=0.075$ (see Fig.~\ref{fig:disk}). Its modal intensity increases linearly with the pump strength, until a second mode of $m_2=7$ turns on at $D_{0,\text{int}}^{(2)}=0.17$, whose non-interacting threshold is at $D_0^{(2)}=0.076$. Beyond $D_{0,\text{int}}^{(2)}$, an anomalous behavior occurs: unlike in a typical laser where the onset of a new mode just slightly slows down the increase of the intensity of the the existing mode(s) [Fig.~\ref{fig:schematics}(c)], the power slope of mode 1 actually becomes negative, and it stops lasing at $D_{0,\text{off}}^{(1)}=0.19$. Note that we have used here full SALT, for which the intensity variation with pump need not be linear \cite{Science}, but in fact it is quite linear as predicted by the SPA-SALT approximation. This phenomenon is confirmed with good agreement by directly solving the differential form of SALT numerically \cite{directMethod}, without the expansion in the CF basis (not shown).

To understand the origin of this mode switching behavior we analyze in more detail the criterion for IMS in the two mode case derived from the SPA-SALT approximation, that $\tilde{S_1} < 0$.
% As shown above, this is a good approximation in our microdisk laser. Not true, spasalt isn't good here quantitatively, only qualitatively.
%In this case the frequency dependence of the CF states can also be neglected.
Let us first discuss the typical behavior in a two-mode laser shown schematically in Fig.~\ref{fig:schematics}(c). One usually finds the cross-interaction coefficients to be much smaller than the self-interaction coefficients, i.e., $\chi_{12},\chi_{21}\ll\chi_{11},\chi_{22}$. Therefore, for two lasing modes with similar non-interacting thresholds, we immediately see that $\tilde{S}_1 \lesssim S_1$ from Eq.~(\ref{eq:S}),
which indicates that the onset of the second lasing mode slightly reduces the power slope of the existing mode, and $\tilde{S_1}$ remains positive.
Consequently, the IMS behavior shown in Fig.~\ref{fig:disk} must be due to a strong cross-interaction.

We now derive the relevant criterion for IMS in terms of the self- and cross-interaction coefficients.
We first note that by definition $D^{(0)}_{2,\text{int}},D^{(2)}_0>D^{(1)}_0$, i.e., the second mode turns on at a higher pump strength than the first mode, with or without modal interactions. Therefore, we find
\be
\frac{\chi_{21}}{\chi_{11}} < \frac{D_0^{(1)}}{D_0^{(2)}} < 1
\ee
from Eq.~(\ref{eq:Dth2}).
With this observation, we derive from \Eq{eq:S} the following criterion for a negative $\tilde{S}_1$:
\be
\frac{\chi_{21}}{\chi_{11}}< \frac{\chi_{22}}{\chi_{12}} < \frac{D_0^{(1)}}{D_0^{(2)}}. \label{eq:ineq}
\ee
When satisfied, mode 1 is switched off at
\be
D_{0,\text{off}}^{(1)} = \frac{1-\frac{\chi_{22}}{\chi_{12}}}{\frac{D_0^{(1)}}{D_0^{(2)}}-\frac{\chi_{22}}{\chi_{12}}}D_0^{(1)}.\label{eq:Doff1}
\ee
This expression is consistent with \Eq{eq:Dth2}, which leads to $D_{0,\text{off}}^{(1)}>D_{0,\text{int}}^{(2)}$, i.e., the termination of mode 1 happens after the onset of mode 2.

The criterion (\ref{eq:ineq}) requires a cross-interaction coefficient larger than the self-interaction coefficient of the ``killer'' mode 2 and smaller than that of the ``victim'' mode 1:
\be
\chi_{22} < \chi_{21}\approx\chi_{12}<\chi_{11}. \label{eq:ineq2}
\ee
To find the circumstances in which these inequalities can hold, we note that in the definition (\ref{eq:chi}) of the interaction coefficient $\chi_{pq}$, the intensity profile of mode $p$ $(q)$ can be treated as the weighting function for mode $q$ $(p)$. To have a large $\chi_{pq}$, the best weighting function would be a delta function at the peak(s) of mode $q$ $(p)$. Therefore, for the criterion (\ref{eq:ineq2}) to hold, the intensity profile of mode 1 needs to be peaked very close to the peak(s) of mode 2, and it should be more localized. For the two whispering-gallery modes shown in Fig.~\ref{fig:disk}, the first mode ($m_1=8$) indeed has a slightly narrower peak in the radial direction that overlaps with the peak of the second mode (see the inset). As a result, we find $\chi_{21}/\chi_{11}=0.977$,  $\chi_{22}/\chi_{12}=0.979$, and \Eq{eq:ineq2} is satisfied.
In addition, we find that the complete criterion (\ref{eq:ineq}) also holds, because $D_0^{(1)}/D_0^{(2)}=0.981$.
%The latter is true due to the material absorption ($\im{n}=0.01$) and the strong spectrum overlap of the second mode with the gain ($|\omega_2-\omega_a|\ll|\omega_1-\omega_a|$); if the radiation were the only loss mechanism and the gain curve were flat, the threshold of mode 1 would be about twice of mode 2.

Due to the small differences between the ratios $\chi_{21}/\chi_{11}$,  $\chi_{22}/\chi_{12}$, and $D_0^{(1)}/D_0^{(2)}$, which appear in the denominators of Eqs.~(\ref{eq:Dth2}) and (\ref{eq:Doff1}), the quantitative values of the onset threshold of mode 2 and the termination threshold of mode 1 as given by SPA-SALT ($D_{0,\text{int}}^{(2)}\approx0.41$, $D_{1,\text{off}}^{(2)}\approx0.70$) are susceptible to the inaccuracy introduced by the single-pole approximation and do {\it not} agree very well with their actual values ($D_{0,\text{int}}^{(2)}\approx0.17$, $D_{1,\text{off}}^{(2)}\approx0.19$).  Nevertheless, the predictive power of Eq.~(\ref{eq:ineq}) can be confirmed by showing the absence of IMS when Eq.~(\ref{eq:ineq}) is violated by some margin.

\begin{figure}[t]
\centering
\includegraphics[clip,width=0.9\linewidth]{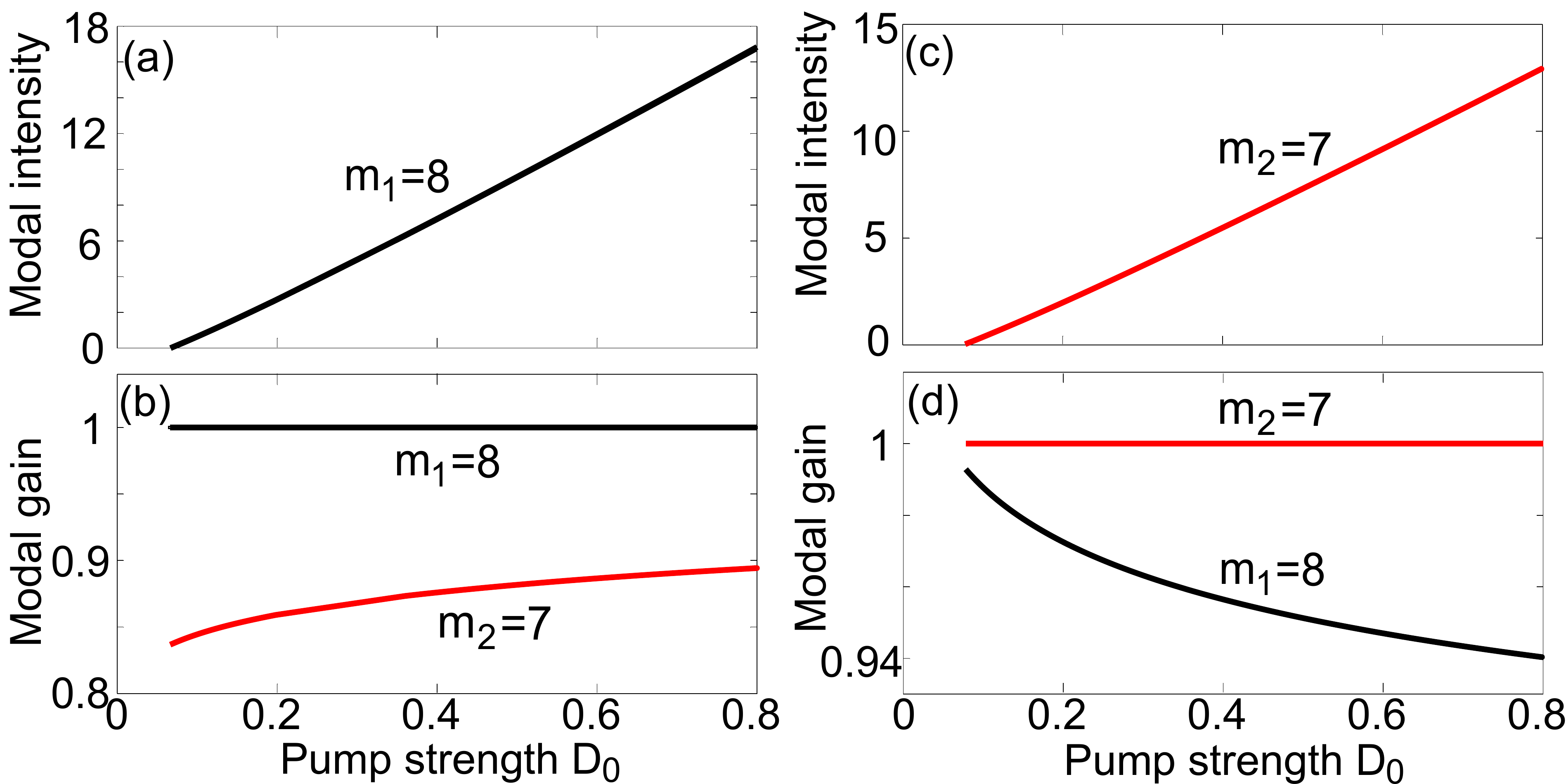}
\caption{(a,c) Same as the modal intensity plot shown in Fig.~\ref{fig:disk} but with $\omega_aR/c=0.50$ (a) and $\omega_aR/c=0.48$ (c). Single-mode lasing is observed and mode switching does not occur in either case, even when the pump power is high above threshold.
(b,d) The modal gain of these two modes in (a,c), confirming that the $m_2=7$ mode in (a) and the $m_1=8$ mode in (c) do not lase. Modal gain is clamped at 1 for lasing modes and stays below 1 for non-lasing modes \cite{SPASALT}.
% In both (a,c) the second mode to turn on above $D_0=0.8$ is a lower $m$ mode which does not satisfy the necessary condition (\ref{eq:ineq2}) and not shown.
}\label{fig:breakdown}
\end{figure}

We can most conveniently violate the SPA-SALT criterion for IMS by varying the last ratio of Eq.~(\ref{eq:ineq}) in our numerical calculation; we achieve this by moving the gain center $\omega_a$.
This approach changes the non-interacting thresholds of these two modes but barely affects their lasing frequencies and mode patterns. Hence $\chi$'s and the first two ratios in Eq.~(\ref{eq:ineq}) can be treated as unchanged.
We first move the gain center closer to the $m_1=8$ mode, which reduces its non-interacting threshold while increasing that of the $m_2=7$ mode. Eventually the last ratio in Eq.~(\ref{eq:ineq}) becomes smaller than the middle one and Eq.~(\ref{eq:ineq}) breaks down. In Fig.~\ref{fig:breakdown}(a) we show such a case with $\omega_a=5.0$, where $D_0^{(1)}=0.0663, D_0^{(2)} = 0.0792$, and $D_0^{(1)}/D_0^{(2)}=0.837<\chi_{21}/\chi_{11},\chi_{22}/\chi_{12}$. Indeed we find that IMS does not take place, and the $m_2=7$ mode is suppressed even when the pump strength is 10 times its non-interacting threshold. Next we move the gain center closer to the $m_2=7$ mode, which becomes the first one to lase. Thus the necessary condition (\ref{eq:ineq2}) breaks down, since now the mode index 1 and 2 are exchanged. Therefore, the criterion (\ref{eq:ineq}) also breaks down, and again we find that IMS does not take place as a result. In Fig.~\ref{fig:breakdown}(c) we show such a case with $\omega_a=4.80$, where the non-interacting thresholds for the $m=7,8$ modes are 0.0762, 0.0767, respectively. As a conclusion, we find that Eq.~(\ref{eq:ineq}) from the SPA-SALT derivation gives a reliable criterion for the occurrence of IMS.

We also note that in Fig.~\ref{fig:disk} we have chosen the gain center to be in close vicinity of the frequency of the second mode. This choice was made such that the non-interacting threshold of the second mode, which has a lower $Q$-factor, is not too high to violate the last inequality of Eq.~(\ref{eq:ineq}). However, IMS in general does not require the gain center to be closer to the second mode than the first mode.

Finally, we address in more detail the difference between the IMS phenomenon studied here and the ``mode-killing" effect found previously in 2D  quasi-ballistic random lasers (QBRL) \cite{Science}.  Because
QBRL have very low $Q$, the lasing modes change considerably in their spatial mode patterns above threshold and their lasing frequencies have strong power dependence, both of which are not included in SPA-SALT. It was found that when two lasing frequencies began to cross as a function of power, one of the solutions would reduce in intensity to zero, but the remaining mode would change dramatically its spatial pattern and would evolve into a fairly equal mixture of the two modes before the frequency crossing.   This ``mode-mixing" scenario requires very low $Q$ and does not occur even in more disordered diffusive random lasers. %\cite{Ge_thesis}.
It cannot play a role in the current example for two reasons.  First, with uniform pumping the microdisk laser conserves the azimuthal quantum number $m$, so our two interacting modes of different $m$'s do not mix spatially through the non-linear interaction; they only affect each other through gain saturation. %(In principle they can mix with other modes of the same $m$ but a different radial quantum number, which are further away in frequency and thus have a very weak effect.)
Second, the frequencies of these two modes are well separated and do not change with increasing pump.
In fact we would not expect IMS to occur in general in random lasers, because we require quite large cross-interaction coefficients, which are not likely to occur between two modes of random lasers.% \cc{away from the localization regime}.

\section{IMS with an evolving pump profile}

In this section we show that when IMS does not occur for a fixed pump profile, it can be facilitated by a pump profile that evolves with the pump strength; IMS can then be induced as a result of the combined actions of a linear effect (i.e., change of non-interacting thresholds with the varied pump profile) and the non-linear modal interaction.

Conceptually IMS is less surprising in this case. First one chooses a system in which different modes are to some extent spatially distinct, so that by changing the spatial profile of the pump one can favor one mode over another with similar $Q$ value. Then one simply defines a ``pump trajectory" \cite{EP} which initially favors one mode and then evolves so as to favor the other. Eventually the effective gain of the second mode becomes so high that it forces the first mode below threshold by gain saturation \cite{bistability2}.

\begin{figure}[htbp]
\centering
\includegraphics[clip,width=0.9\linewidth]{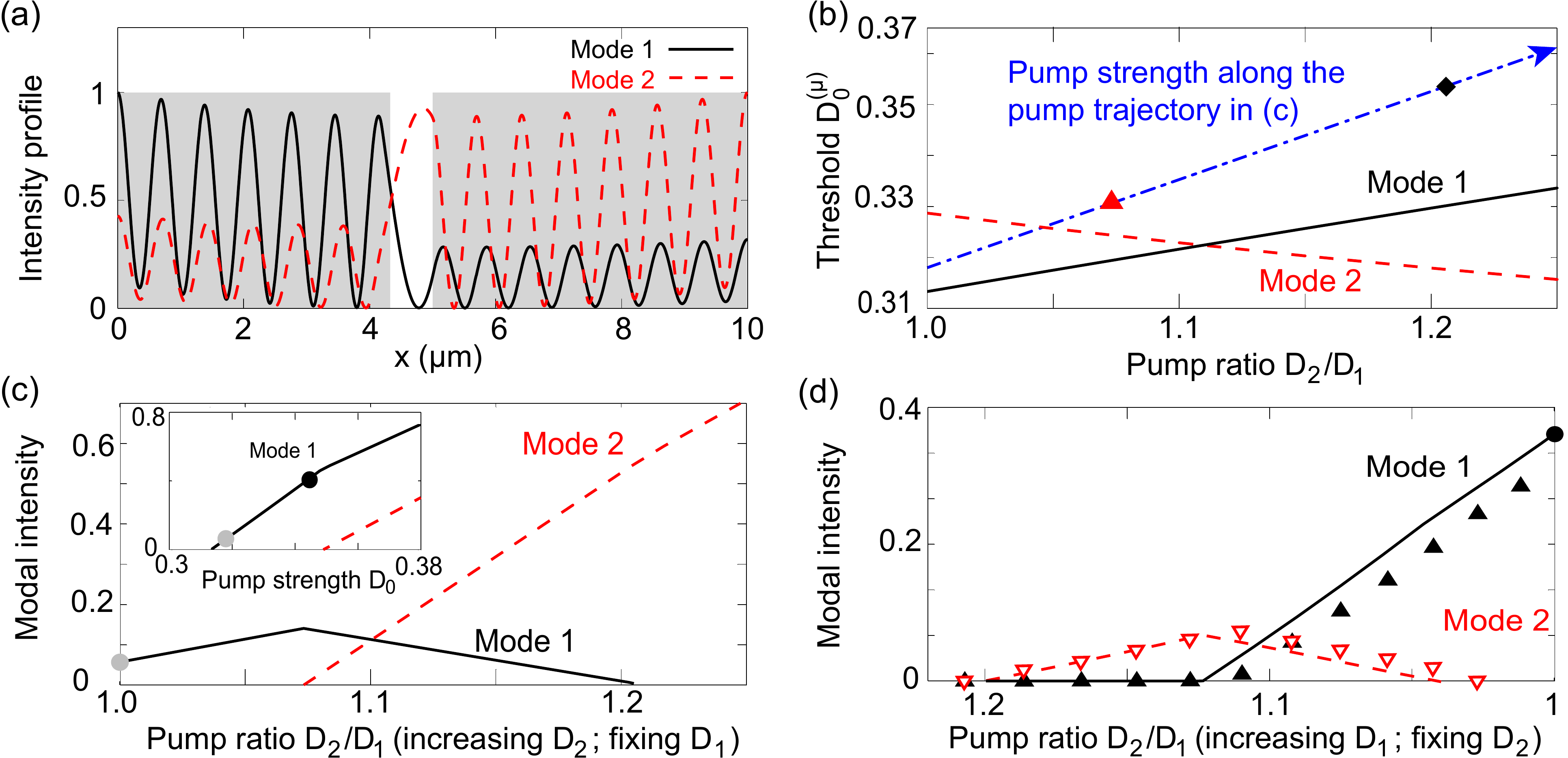}
\caption{Interaction-induced mode switching facilitated by an evolving pumping in two coupled 1D cavities of index $n=3$. Their lengths are $L_1=4.2\mu{m}$, $L_2=5.0\mu{m}$ and the gap width between them is $W = 0.8\mu{m}$. $L\equiv L_1+L_2+W=10\mu{m}$. (a) Intensity profiles of two modes with wavelength $\lambda_1=4.15\mu{m}$, $\lambda_2=4.28\mu{m}$. Shaded areas show the two cavities.
(b) Their non-interacting thresholds (solid and dashed lines) versus the pump ratio $D_2/D_1$. The average pump strength $D_0$ along the pump trajectory used in (c) is shown as the dash-dotted line, and the arrow indicates the direction of increasing $D_1$. The triangle and diamond on it mark the onset of mode 2 and the termination of mode 1. The gain curve is centered at  $\omega_aL/c=15$ ($\lambda_a = 4.19\mu{m}$) and its width is $\gper L/c=1$.
(c) IMS of mode 1 when the pump strength in the right cavity is increased while the pump strength in the left cavity is fixed at $D_1=0.318$. Inset: Mode switching does not occur for uniform pumping ($D_1=D_2$).
(d) IMS of mode 2 when the pump strength in the left cavity is increased while the pump strength in the right cavity is fixed at $D_2=0.344$. Good agreement between SALT (lines) and FDTD simulations (triangles) are shown. $\gamma_\|/\gper=5\times10^{-4}$ is used in FDTD simulations.}\label{fig:coupled}
\end{figure}

However, this intuitive picture does not always hold, because there is no guarantee that the effective gain of the first mode will reduce eventually as the overall pump power increases (see the quantitative discussion below Eq.~(\ref{eq:ineq3}) near the end of this section). More importantly, evolving the pump profile is an important generalization because one relies much less on fine tuning of parameters of the cavity and gain medium to generate IMS, which then occurs more robustly.
%And we shall see, the generalization of Eq.~(\ref{eq:ineq}) in this case provides quantitatively accurate predictions for IMS.
This is exemplified in Fig.~\ref{fig:coupled} using two coupled 1D laser cavities with slightly different lengths ($L_1<L_2$) and independent pump controls ($D_1,D_2$). The gap between the two cavities ($W$) is much shorter than their lengths, and there exist pairs of lasing modes that have similar frequencies but reside asymmetrically in these two cavities.

One such pair (mode 1 and 2) are shown in Fig.~\ref{fig:coupled}(a).
%The asymmetry of their intensity profiles is created by the different cavity lengths, which in turn leads to a different dependence on the asymmetry of the pump profile.
Mode 1 has a stronger intensity in the left cavity, while mode 2 does the opposite. If the two coupled cavities are pumped uniformly, we find that mode 1 turns on first at $D_0=0.313$, and its intensity continues to increase after mode 2 turns on at $D_0=0.349$ [see Fig.~\ref{fig:coupled}(c); inset]. The absence of IMS in this case is well captured by the criterion (\ref{eq:ineq}), which does not hold since ${\chi_{22}}/{\chi_{12}}=1.64$ is larger than ${D_0^{(1)}}/{D_0^{(2)}}=0.953$.

To generate IMS for this system and these two modes, we evolve the pump profile as we increase its average strength, by uniformly pumping each cavity but with different weights. Specifically, the pump profile $f(x)$ is normalized by
$\int_{L_1\cup L_2} f(x) dx = f_1L_1+f_2L_2\equiv L_1+L_2$, and the average pump strength is given by $D_0 = (D_1L_1+D_2L_2)/(L_1+L_2)$, with $D_{1(2)}\equiv D_0f_{1(2)}$ by definition. Since mode 1 is stronger in the left cavity, its threshold increases if we pump the right cavity preferably [see Fig.~\ref{fig:coupled}(b)]; the opposite holds for mode 2, because it has a stronger intensity in the right cavity. For a fixed ratio $f_2/f_1<1.11$ (including the uniform pumping case), mode 1 is the first mode to lase when the pump strength is increased.

Starting from uniform pumping and $D_1=D_2=0.318$ (grey dot in the inset of Fig.~\ref{fig:coupled}(c)), we increase $D_2$ while keeping $D_1$ fixed. The modal intensity of mode 1 continues to increase until the onset of mode 2 at $D_2=0.342$ ($f_2/f_1=1.07$), after which it reduces with $D_2$ [Fig.~\ref{fig:coupled}(c)]. Mode 1 is switched off eventually at $D_2=0.385$ ($f_2/f_1=1.21$). It is clear that the non-linear modal interaction still plays a crucial role in this mode switching scenario; mode 1 would have continued to lase above $D_2=0.385$ if not interacting with mode 2, because the average pump strength is still above its non-interacting threshold [Fig.~\ref{fig:coupled}(b)]. This behavior is distinct from the self-termination of a lasing mode due to an exceptional point as reported in \cite{EP,EP_exp, EP_CMT}; the latter is a linear effect and does not depend on the non-linear modal interactions.

%We note that mode mixing is again nonessential in the IMS demonstrated here. (Not nec to address this again).

Interestingly, with this additional control knob, the IMS scenario can be interchanged if a different initial pump configuration is chosen. In Fig.~\ref{fig:coupled}(d) we start with $D_1=0.287$ and $D_2=0.344$ (%$D_0=0.318$ and
$D_2/D_1=1.2$). Only mode 2 lases since the pump strength is just above its non-interacting threshold but below that of mode 1. As we increase $D_1$ while keeping $D_2$ fixed, the intensity of mode 2 increases linearly until $D_1>0.306$ ($D_2/D_1<1.12$), beyond which mode 1 turns on and switches off mode 2 at $D_1=0.331$ ($D_2/D_1=1.04$).
% This observation also implies that if we reverse the process, i.e., by reducing the pump in the left cavity starting with $D_1>0.331$, we will see mode 2 actually becomes lasing until $D_1$ is too small.

For this one-dimensional example it is relatively easy to provide a test of our results independent of SALT by using time integration of the full Maxwell-Bloch equations \cite{OpEx08,N_level,directMethod}.  As shown in Fig.~\ref{fig:coupled}(d), such calculation also predicts IMS and agrees with SALT, finding less than a $1\%$ difference between the values of the pump strength at which mode 1 is killed. This difference is within the typical range (a few percent) when comparing SALT and FDTD results \cite{OpEx08,N_level,directMethod}.
We also note that our coupled-cavity laser does not show signs of bi-stability \cite{Lamb,bistability1,bistability2,bistability3}, i.e., the frequencies and intensities of the lasing modes are unique for a given pump combination $D_1$ and $D_2$, even though it can be reached by evolving different initial pump configurations. One such comparison is shown by the grey dots in Fig.~\ref{fig:coupled}(c) and its inset, which give the same intensity of mode 1 at $D_1=D_2=0.318$. Another comparison is shown by the black dots in Fig.~\ref{fig:coupled}(d) and the inset of Fig.~\ref{fig:coupled}(c), which give the same intensity of mode 1 at $D_1=D_2=0.344$. The absence of bistability here is due to the conditions $\gper,\Delta\gg\gamma_\|$ we consider, with which a stationary inversion can be assumed in the two-mode regime \cite{TS,OpEx08,directMethod,HS2}; it is exact in the single mode regime.

\begin{figure}[htbp]
\centering
\includegraphics[clip,width=0.7\linewidth]{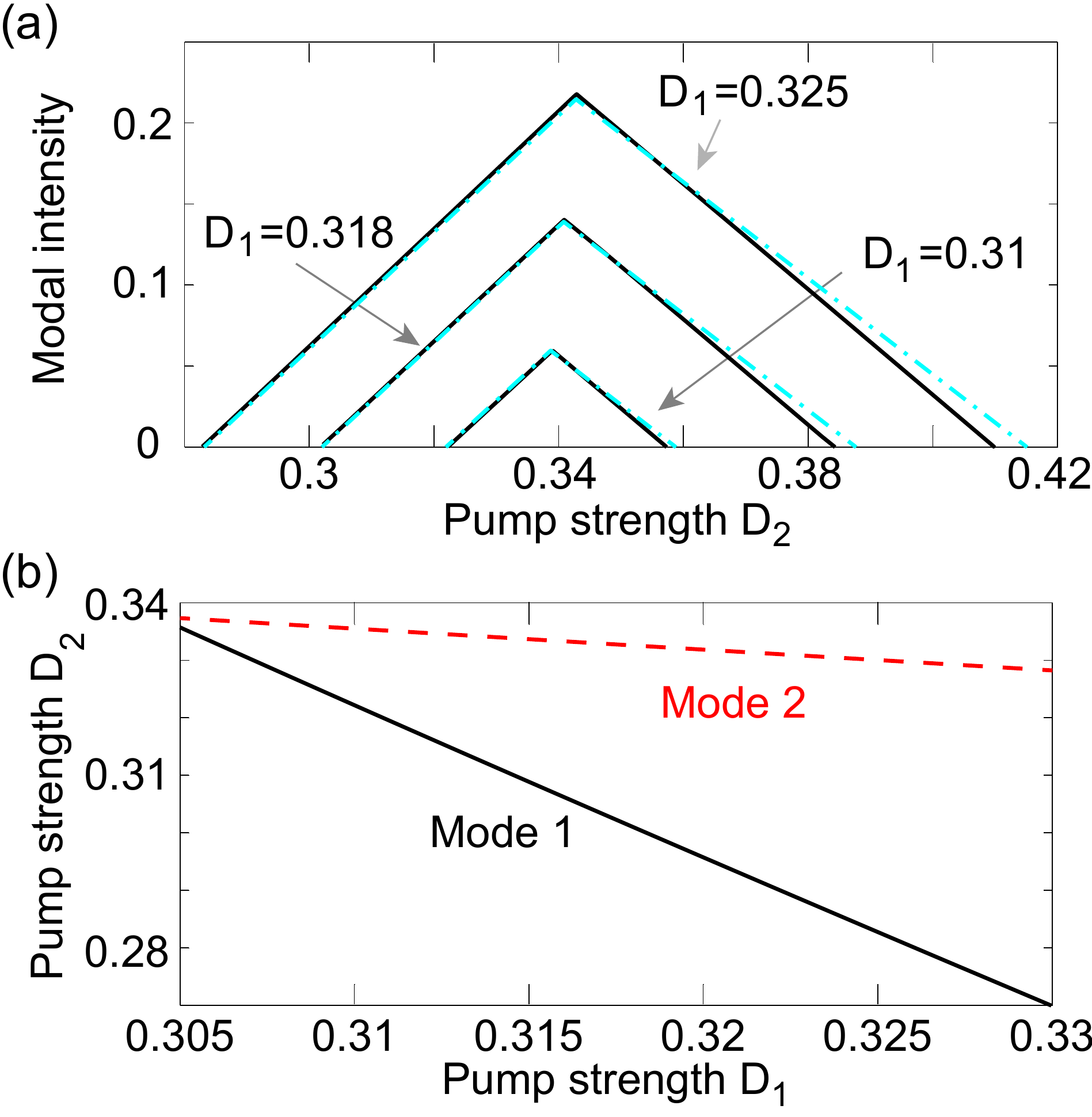}
\caption{Invariant shape of the IMS triangle. (a) Same as Fig.~\ref{fig:coupled}(c) but with different values of the fixed pump strength $D_1$. Modal intensity of mode 2 is not shown for clarity. Solid lines show the full SALT results and dash-dotted lines show the SPA-SALT approximations [Eq.~(\ref{eq:SPASALT_coupled})]. (b) Non-interacting thresholds in terms of $D_2$ when $D_1$ is varied.}\label{fig:coupled2}
\end{figure}

Besides the flexibility of exchanging the killer mode and the victim mode, IMS is also robust and exists in a wide parameter space when evolving the pump profile. For example, IMS shown in Fig.~\ref{fig:coupled}(c) does not rely on the specific value of the fixed pump strength $D_1$; it always occurs as $D_2$ is increased, provided that the mode 1 still lases first. Surprisingly, we find that the modal intensity $I_1$ along these different pump trajectories are all invariant in shape as a function of $D_2$, as shown in Fig.~\ref{fig:coupled2}(a): the triangular area underneath the modal intensity (``IMS triangle") is simply scaled and shifted \cite{bibnote:2}.

To understand this observation, we first compare the thresholds (in terms of $D_2$) of both modes when evolving the pump profile from different $D_1$ values. As we show in Fig.~\ref{fig:coupled2}(b), if we increase the $D_1$ value we start with, the non-interacting threshold of mode 1 (in terms of $D_2$) is reduced more than that of mode 2, thanks to the stronger intensity of mode 1 in the left cavity. Therefore, we expect mode 1 to lase in a wider range of $D_2$, which enlarges the IMS triangle. To understand the shape invariance of the IMS triangle, we note that it is equivalent to show that the power slope of mode 1 with respect to $D_2$ does not depend on the fixed $D_1$ value along each pump trajectory, both before and after the onset of the second mode.

We resort to SPA-SALT again to explain this property, which also gives the IMS criterion when two independent pumps are applied. The constrained two-mode equations (\ref{eq:SPASALT}) now become:
\be
M
\begin{pmatrix}
I_1 \\ I_2
\end{pmatrix}
=
P
\begin{pmatrix}
D_1 \\ D_2
\end{pmatrix}
-
\begin{pmatrix}
1 \\ 1
\end{pmatrix},
\quad
P
\equiv
\begin{pmatrix}
\frac{W_{11}}{D_0^{(1)}} & \frac{W_{12}}{D_0^{(1)}} \vspace{4pt} \\
\frac{W_{21}}{D_0^{(2)}} & \frac{W_{22}}{D_0^{(2)}}
\end{pmatrix},\label{eq:SPASALT_coupled}
\ee
where we have used the threshold CF states \cite{SPASALT} that have the property that one such state is the threshold solution, even for a non-uniform index cavity and non-uniform pumping. $M$ here is the same as given in Eq.~(\ref{eq:SPASALT}), $D_0^{(\mu)}$ is the non-interacting threshold of mode $\mu$ with $f_1=f_2$, and
\be
W_{\mu\nu} = \frac{1}{V}\left|\int_\text{cavity} dx\,u_\mu(x) \bar{u}_\mu^*(x)\,\eta_\nu(x)\right| \label{eq:W}
\ee
is the overlapping factor of mode $\mu$ (its corresponding CF state $u_\mu(x)$ to be exact) with the pump in cavity $\nu$ $(\nu=1,2)$, where $\eta_\nu(x)=1$ in cavity $\nu$ and 0 elsewhere in our example above. $W_{\mu\nu}$ represents the linear effect we mentioned at the beginning of this section, which indicates that the evolving pump profile changes the non-interacting thresholds of the lasing modes. For example, $D_0^{(1)}/W_{12}$ and $D_0^{(2)}/W_{22}$, whose inverse appear in the matrix $P$ in Eq.~(\ref{eq:SPASALT_coupled}), are approximations of the non-interacting thresholds of mode 1 and 2 when pumping only cavity 2 \cite{ge_NP}. %Similar to the discussion of the interactions coefficients, $W_{\mu\nu}$ can be treated as real for high-$Q$ modes.
The SPA-SALT equations (\ref{eq:SPASALT_coupled}) give good approximations of the modal intensity of interacting thresholds as shown in Fig.~\ref{fig:coupled2}(a).

Assuming a fixed pump strength $D_1$ and varying $D_2$ as in Figs.~\ref{fig:coupled}(c) and \ref{fig:coupled2}(a), we immediately find that the power slope of mode 1 with respect to $D_2$, i.e., $S_1=W_{12}/[\Gamma_1\chi_{11}D_0^{(1)}]$, is independent of the fixed pump strength $D_1$ in cavity 1. Likewise, after the onset of mode 2, the power slope becomes
\be
\tilde{S}_1 =
\frac{\frac{\chi_{22}}{\chi_{12}}-\frac{D_0^{(1)}W_{22}}{D_0^{(2)}W_{12}}}{\frac{\chi_{22}}{\chi_{12}} - \frac{\chi_{21}}{\chi_{11}}} S_1, \label{eq:S_coupled}
\ee
which is also independent of the fixed $D_1$ value along each pump trajectory. This concludes our proof of the shape invariance of the IMS triangle.

The factor $W_{22}/W_{12}$ in the numerator of Eq.~(\ref{eq:S_coupled}) makes IMS rely much less on fine tuning of parameters of the cavity and gain medium. It is basically an indicator that shows how much mode 2 (the killer mode here) benefits more from the second pump $D_2$ than mode 1 (the victim mode). For uniform pumping this factor is 1 by definition [see  Eqs.~(\ref{eq:biortho}) and (\ref{eq:W})] and we recover Eq.~(\ref{eq:S}). In contrast to the uniform pumping case, here even with $\chi_{11},\chi_{22}$ larger than $\chi_{12},\chi_{21}$ as in a typical laser (which gives a positive denominator in Eq.~(\ref{eq:S_coupled})), a negative power slope $\tilde{S}_1$ and hence IMS can still be induced by simply focusing the second pump onto mode 2 and avoiding mode 1, such that
\be
\frac{W_{22}}{W_{12}}>\frac{D_0^{(2)}\chi_{22}}{D_0^{(1)}\chi_{12}}.\label{eq:ineq3}
\ee
As we mentioned previously, ${\chi_{22}}/{\chi_{12}}=1.64$ is larger than ${D_0^{(1)}}/{D_0^{(2)}}=0.953$ in the example shown in Figs.~\ref{fig:coupled} and \ref{fig:coupled2}, and the criterion (\ref{eq:ineq}) for IMS with uniform pumping does not hold. However, $W_{22}/W_{12}$ is 2.64 in this case and the criterion (\ref{eq:ineq3}) holds, thanks to the asymmetric intensity patterns of mode 1 and 2 in the two cavities [Fig.~\ref{fig:coupled}(a)]. Equation~(\ref{eq:ineq3}) also points out the occurrence of IMS by evolving pump profile is more subtle than the intuitive picture discussed at the beginning of this section: if the second pump favors mode 2 over mode 1(i.e., ${W_{22}}/{W_{12}}>1$) but fails to meet the quantitative requirement imposed by Eq.~(\ref{eq:ineq3}), IMS does not occur either.

Last but last least, the lasing modes for two \textit{identical} cavities near threshold are just the symmetric and antisymmetric superpositions of the same individual cavity mode in these two cavities, and their intensity profiles are both {\it symmetric} in the two cavities, resulting in $W_{22}/W_{12}\approx1$ and the absence of IMS, even with independent controls of $D_1$ and $D_2$. One exception is when the system is so lossy that the cavity decay rate becomes larger than the coupling between the two cavities, which can result in asymmetric intensity profiles of the two modes in these two cavities (see the discussion of the ``broken symmetry phase" in \cite{EP_CMT}).

\section{Conclusion}

We have discussed a mode switching behavior that occurs in lasers with strong non-linear modal interactions. In this process the identities of the lasing modes are preserved, including their spatial patterns and frequencies.
IMS is a robust phenomenon and can in principle occur in any type of laser, including standard systems such as a microdisk and coupled one-dimensional cavities.
%This indicates that mode mixing does not play a role here, which was attributed to a seemingly similar effect in random lasers \cite{Science}. Let's not keep bringing this up.
In IMS the power slope of a lasing mode turns negative when another lasing mode turns on, resulting in the switching off of the first mode in a linear fashion. Qualitatively this phenomenon is due to strong cross-saturation effects, and it is explained by the satisfaction of two inequalities involving the modal interaction coefficients and the non-interacting thresholds of the modes. Specifically, using the SPA-SALT approximation, we have identified a criterion for IMS with a fixed pump profile, which requires the cross-interaction coefficient to be stronger than the self-interaction coefficient of the killer mode but smaller than that of the victim mode, as well as a constraint on their non-interacting thresholds.
% More importantly, in \cite{Science} it is the newly emerged mode 2 that is switched off by interacting with an existing mode 1, leading to a lasing sequence of $1\rightarrow(1,\,2)\rightarrow$1 as the pump power is increased. % which cannot be used as a switch.
%Here however, the opposite takes place with the existing mode 1 being switched off, leading to a lasing sequence of 1$\rightarrow$(1,\,2)$\rightarrow$2.
These conditions can be satisfied, for example, when the two modes have similar peak structures and the killer mode is more localized. More importantly from the point of view of controlling the mode spectrum of lasers, even when these conditions are not satisfied for a fixed spatial profile of the pump, IMS can be generated by evolving the spatial pattern of the pump as the total pump power increases, no matter whether the laser system consists of a single cavity or coupled cavities.
This mode switching behavior seems to occur in exciton-polariton condensates as well \cite{GPE}. IMS may have potential applications in robust and flexible all-optical switching \cite{Kroha}, stepwise-tunable single-mode laser sources, and optical memory cells \cite{memory}.
An existing experimental setup for its demonstration may consist of a microdisk cavity and a spatial light modulator for the optical pump, which has been employed successfully to demonstrate pump-controlled emission directionality \cite{Seng_APL14} and modal interactions \cite{Seng_PRA15}.

\section*{Acknowledgments}
We thank Konstantinos Makris for helpful discussions. L.G. acknowledges partial support by PSC-CUNY Grant No. 68698-0046 and NSF Grant No. DMR-1506987. D.L. and S.G.J. acknowledges partial support by the AFOSR MURI grant No. FA9550-09-1-0704 and by the Army Research Office through the Institute for Soldier Nanotechnologies (ISN), Grant No. W911NF-07-D-0004. A.C. and A.D.S. acknowledge support under NSF Grant No. DMR-1307632. S.R. acknowledges support by the Vienna Science and Technology Fund (WWTF) through Project No. MA09-030, and by the Austrian Science Fund (FWF) through Projects No. F25-P14 (SFB IR-ON), No. F49-P10 (SFB NextLite). H.C. acknowledges support by NSF Grant No. DMR-1205307. H.E.T. acknowledges NSF Grant No. EEC-0540832 (MIRTHE) and DARPA Grant No. N66001-11-1-4162.


\begin{thebibliography}{99}
\bibitem{bibnote:2}The same behavior happens to $I_2$ in Fig.~\ref{fig:coupled}(d) when different initial values of $D_2$ are chosen.
\end{thebibliography}
\end{document}